\documentclass[prd,amsfonts,nofootinbib]{revtex4}
\def\321{$\mathrm{SU(3) \otimes SU(2)_L \otimes U(1)}$ }
\def\422{$\mathrm{SU(4) \otimes SU(2)_L \otimes SU(2)_R}$ }
\def\vev#1{\left\langle #1\right\rangle}
\usepackage[pdftex]{graphicx}  
\usepackage{color}
 \usepackage[english]{babel}

\begin{document}

\title{\bf Quark-Lepton Mass Relation in a Realistic
  $A_4$ Extension of the Standard Model} \author{ {S.~F.~King}$^a$,
  {S.~Morisi}$^b$, {E.~Peinado}$^c$,
  {J.~W.~F.~Valle}$^d$, \\
  $^a${\small \it School of Physics and Astronomy, University of
    Southampton,
    Southampton, SO17 1BJ, U.K. }\\
  $^b${\small \it Institut f{\"u}r Theoretische Physik und Astrophysik,}\\
  {\small \it Universit{\"a}t W{\"u}rzburg, 97074 W{\"u}rzburg, Germany}\\
  $^c${\small \it INFN, Laboratori Nazionali di Frascati,}\\
  {\small \it Via Enrico Fermi 40, I-00044 Frascati, Italy}\\
  $^d${\small \it AHEP Group, Institut de F\'{\i}sica Corpuscular - C.S.I.C./Universitat de Val{\`e}ncia}\\
  {\small \it Edificio Institutos de Paterna, Apt 22085, E--46071
    Valencia, Spain} }

\begin{abstract}

  We propose a realistic $A_4$ extension of the Standard Model 
  involving a particular quark-lepton mass relation, namely that 
  the ratio of the third family mass to the geometric mean of the 
  first and second family masses are equal for down-type quarks 
  and charged leptons. This relation, which is approximately 
  renormalization group invariant, is usually regarded as arising 
  from the Georgi-Jarlskog relations, but in the present model 
  there is no unification group or supersymmetry. In the neutrino 
  sector we propose a simple modification of the so called 
  Zee-Wolfenstein mass matrix pattern which allows an acceptable 
  reactor angle along with a deviation of the atmospheric and solar 
  angles from their bi-maximal values. Quark masses, mixing angles 
  and CP violation are well described by a numerical fit. 

\end{abstract}
\maketitle

\section{Introduction}

Supersymmetric Grand Unified Theories (SUSY GUTs) are very attractive
from the theoretical point of view as they allow to obtain the SM from
a single unified gauge group~\cite{Georgi:1974sy,pati:1974yy}. Apart
from predicting for instance the quantization of electric charge, they
reduce the number of free parameters. For example, they give the right
value for the electroweak mixing angle and may provide a good
framework for the understanding of the flavour problem. Indeed, several
GUT models have been studied in the literature, having as prediction a
mass relation between down-quark masses and the charged leptons. For
instance in the $SU(5)$ unified framework Georgi and Jarlskog have
found the mass relation~\cite{Georgi:1979df}
\begin{equation}\label{gut}
m_e=\frac{1}{3}m_d\,,\qquad m_\mu=3m_s\,,\qquad m_\tau=m_b\,,
\end{equation}
which is in good agreement with data to first approximation, assuming
that holds at the GUT scale, and taking into account renormalization
group running to low energies, with suitable SUSY threshold effects.
Such mass relations are very welcome since, by itself, the Standard
Model sheds no light on the flavour problem.

However, the Large Hadron Collider has so far not found any evidence
of physics beyond the Standard Model (SM).  Indeed, the only major
discovery to date has been that of a new boson which is entirely
consistent with the properties of the SM Higgs boson, arising from a
single Higgs doublet $H$.  In particular the LHC has not so far found
any evidence for Supersymmetry (SUSY) as indicated within the simplest
Grand Unified Theories (GUTs), namely those which do not involve an
intermediate scale such as $SU(5)$.

Here we advocate an alternative TeV-scale approach to the flavour
problem employing just the Standard Model gauge symmetry, supplemented
only by a non-Abelian discrete flavour symmetry.
For the latter we adopt $A_4$, the discrete group of even permutations
of four objects isomorphic to the group of symmetries of the
tetrahedron. It is the smallest group containing triplet irreducible
representations.  Several $A_4$-based flavour models have been
suggested~\cite{babu:2002dz,altarelli:2005yp,Holthausen:2012wz}, for
reviews see Refs.~\cite{Ishimori:2008uc,Hirsch:2012ym,Morisi:2012fg,King:2013eh}.
Recently three of us have proposed an \321 model~\cite{Morisi:2011pt}
based on the discrete family symmetry $A_4$ leading to the
quark-lepton mass relation:
\begin{equation}\label{massrel}
\frac{m_{\tau}}{\sqrt{m_{e}m_{\mu}}}\approx \frac{m_{b}}{\sqrt{m_{s}m_{d}}} \,. 
\label{massrelation}\end{equation}
It is clear that Eq.\,(\ref{massrel}) provides an interesting
generalization of Eq.\,(\ref{gut}) which is found to be in very good
agreement with data. Given that it is approximately renormalization
group invariant, it holds at all mass scales~\cite{Morisi:2011pt}. In
contrast to the Georgi-Jarlskog relation, Eq.~(\ref{massrelation})
arises just from the flavour structure of the model, and the fact from
the existence of two Higgs doublets selectively coupled to the up- and
down-type fermions~\footnote{Such structure is required in
  supersymmetric models, but the mechanism proposed in
  Ref.~\cite{Morisi:2011pt} leading to Eq.~(\ref{massrelation}) is
  more general, relying only of the two-doublet nature of the Higgs
  sector, as mentioned above. Here we abandon the use of
  supersymmetry. }.  Both of these were assigned to be $A_4$ triplets,
with one of them coupling to the down-type quarks and charged
leptons. Note that the model in \cite{Morisi:2011pt} employed an
extended Higgs sector with three families of supersymmetric Higgs
doublets $H_u$ and $H_d$ for which there is presently no evidence,
with present data being consistent with a single Higgs
doublet. Moreover, the model predicted $V_{ub}=0=V_{cb}$. While
providing a good starting point for the CKM matrix, a derivation of
the full quark mixing was lacking.

In the present paper we propose an alternative discrete family
symmetry $A_4$ model, keeping the same motivation for introducing
$A_4$ into the Standard Model~\cite{Morisi:2011pt}, namely, to shed
light on the flavour problem. 
Indeed the model presented here provides a fully realistic description
of all quark and lepton masses and mixing angles, and in particular
reproduces the successful quark-lepton mass relation in
Eq.(\ref{massrel}). In contrast to the previous construction in
\cite{Morisi:2011pt} its remains closer in spirit to the SM, since we
do not assume supersymmetry nor unification, keeping a single Higgs
doublet $H$ instead of multi-Higgs doublets.  In particular, we assign
right-handed up quarks to singlet representation of $A_4$ instead of
triplet as in the original model.
The $A_4$ flavour symmetry is broken by SM singlet flavons which
distinguish up-type quarks from down-type quarks and charged leptons,
with additional flavons in the neutrino sector, where we require extra
Abelian discrete groups to distinguish these sectors.  
Assuming full explicit breaking of $A_4$ through suitable scalar
flavon multiplets we show that this simple modification of the model
in \cite{Morisi:2011pt} can describe all the CKM mixing parameters.
Moreover, it is straightforward to obtain also the so called {\it
  Zee-Wolfenstein} mass matrix pattern in the neutrino sector using
$A_4$ invariance~\cite{Grimus:2008tm}.  However, since it predicts
bi-maximal mixing, current neutrino oscillation data
analysis~\cite{Tortola:2012te} rule out such a pattern
\cite{Jarlskog:1998uf}.  We propose a simple modification of the
Zee-Wolfenstein model where all the mixing angle, as well as the
reactor angle can be reproduced.

\vskip3.mm In the next section we introduce our model, in section
\ref{sec3} we obtain our quark-lepton mass relation, in section
\ref{sec4} we give the fit for the quark mixing parameters, in section
\ref{sec5} we describe the neutrino mass generation mechanism and
study its phenomenological implications, while in section \ref{sec6}
we give our conclusions.

\section{The model}
\label{sec:model}

The matter content and the flavour group assignment are given in
Table \ref{tablamodel}.  Note that all the fermions, apart of the $u_R$
fields, are assigned to triplets of $A_4$~\footnote{Therefore the
  present model can not be embedded in any grand-unified framework.}.
In the scalar sector we have one SM Higgs doublet and four flavon
fields.  With respect to the model of Ref.\,\cite{Morisi:2011pt} we
have extra Abelian symmetries, namely $Z_2^u$, $Z_2^d$ and $Z_3$. The
reason for imposing such a symmetries is because our present model is
not supersymmetric. We are replacing the SUSY-Higgs doublets $H^u$ and
$H^d$ (triplet of $A_4$ in Ref.\,\cite{Morisi:2011pt}) with scalar
$SU_L(2)$-singlets (flavons) triplets of $A_4$ times the standard
model Higgs doublet, namely
\begin{equation}
H^d \,\to\, H\, \varphi_d\,,\quad H^u \,\to\, \tilde{H}\, \varphi_u
\end{equation}
where $\tilde{H}=i\sigma_2 H^*$. It is clear that the $Z_2^u$ and
$Z_2^d$ symmetries glue the $\varphi_u$ and $\varphi_d$ flavons fields
to the up and down quark sectors, respectively, while the extra
$Z_3^{\nu}$ symmetry is used to separate the charged and neutral
fermion sectors.
\begin{table}[h!]
\begin{center}
\begin{tabular}{|c|c|c|c|c|c|c|c||c|c|c|c|c|}
\hline
 & $L$& $l_R$& $Q$ & $d_R$& $u_{R_1}$& $u_{R_2}$& $u_{R_3}$ & $H$ & $\varphi_u$& $\varphi_d$& $\varphi_\nu$&$\xi_\nu$\\
\hline
$A_4$ & 3&3&3&3&$1$&$1''$&$1'$&1&3&3&3&1\\
\hline
$Z_2^u$ & $+$& $+$& $+$& $+$& $-$& $-$& $-$& $+$& $-$& $+$& $+$& $+$\\
$Z_2^d$ & $+$& $-$& $+$& $-$& $+$& $+$& $+$& $+$& $+$& $-$& $+$& $+$\\
$Z_3^{\nu}$ & $\omega$& $\omega^2$& $1$& $1$&$1$&$1$& $1$&$1$&$1$&$1$&$\omega$ &$\omega$ \\
\hline
\end{tabular}\caption{Matter content of the model.}\label{tablamodel}
\end{center}
\end{table}

The Lagrangian for quarks and charged leptons in our model is given by
\begin{eqnarray}
\mathcal{L} &=& \frac{y^d_{\alpha \alpha'}}{M}(Q\,d_R)_\alpha H \varphi_{d_{\alpha'}}+
\frac{y^l_{\alpha\alpha'}}{M}(L\,l_R)_\alpha H \varphi_{d_{\alpha'}}+ 
\frac{y^u_\beta}{M}(Q\,\varphi_u)_\beta \tilde{H} {u_R}_{\beta^\prime} +H.c.,\label{lagrangian}
\end{eqnarray}
where $\alpha,~\alpha^\prime$ label $A_4$ triplets. We remind that the
product of two $A_4$ triplets is given by $3\times 3 = 1+1'+1''+3+3$
where the two triplet contractions can be written as the symmetric and
the antisymmetric ones and denoted as $3_1,\,3_2$\,\footnote{In $A_4$
  there is only one triplet irreducible representation, here $3_1$ and
  $3_2$ are not different irreducible representations, but simply a
  way to indicate different contractions.}.  Thus we have that
$\alpha= 3_1,\,3_2$ while $\alpha\prime=3$, implying that
$y^{d,l}_{\alpha\alpha'}$ gives only two couplings $y^{d,l}_1\equiv
y^{d,l}_{3_1\,3}$ and $y^{d,l}_2\equiv y^{d,l}_{3_2\,3}$.  On the
other hand $\beta$ and $\beta^{\prime}$ can be $1,\,1',\,1''$ in such
a way that $\beta \times \beta^{\prime}=1$. Note that, while the $A_4$
flavour symmetry holds in the (non-renormalizable) Yukawa terms
leading to charged fermion masses, we assume it to be completely
broken in the scalar potential. Indeed, we assume that the scalar
flavon multiplets get vacuum expectation values (vevs) in an arbitrary
direction of $A_4$, preserving none of its subgroups. This can be
easily achieved just by including terms in the scalar potential which
are $SO(3)$ invariant as discussed in \cite{King:2006np}. In this case
the flavon scalar fields get vevs in arbitrary directions of $A_4$,
that is
\begin{equation}
\vev{\varphi_{f}} 
\propto (v_1^f,v_2^f,v_3^f) 
\end{equation}
where $v_1^f\ne v_2^f \ne v_3^f$ and $f=u,d,\nu$. To complete the
model we need also to specify the mechanism of neutrino mass
generation, see Sec.~\ref{sec5}, below.

\section{The charged lepton-quark mass relation}
\label{sec3}

From the $A_4$ contraction rules (see appendix \ref{a4group}) and the
fact that the charged leptons and down-type quarks are in the same
$A_4$ representations, one sees that the charged leptons and down-type
quark mass matrices have the form
\begin{equation}
M_{f}=
\left(
\begin{array}{ccc}
0 & y_{1}^f v_3^f  & y_{2}^f v_2^f \\
y_{2}^f v_3^f & 0 & y_{1}^f v_1^f \\
y_{1}^f v_2^f & y_{2}^f v_1^f & 0
\end{array}
\right),
\label{Mf}
\end{equation}
where $f=d,l$. This special form is the same as obtained in
Ref.~\cite{Morisi:2011pt,Morisi:2009sc}. With the redefinition of
variables:
\begin{eqnarray}
y_{1}^f=a^f/v_2^f\,,\quad
y_{2}^f=b^f/v_2^f\,,\quad
\alpha^f=v_3^f/v_2^f\,,\quad r^f=v_1^f/v_2^f\,,
\end{eqnarray}
the mass matrix for the mass matrix in Eq. (\ref{Mf}) takes the form
\begin{equation}
M_{f}=
\left(
\begin{array}{ccc}
0 & a^f \alpha^f  & b^f  \\
b^f\alpha^f  & 0 & a^f r^f \\
a^f  & b^f r^f & 0
\end{array}
\right)\,.
\label{Mf2}
\end{equation}
Let us now consider the system given by the following three invariants 
\begin{eqnarray}
\det S^f &=& (m^f_1m^f_2m^f_3)^2\label{i1}\\
\mbox{Tr}\, S^f &=& {m^f_1}^2 +{m^f_2}^2 +{m^f_3}^2\label{i2}\\
(\mbox{Tr}\, S^f)^2-\mbox{Tr}\, {S^f}S^f &=& ({m_1^f}^2+{m_2^f}^2){m_3^f}^2+{m_1^f}^2{m_2^f}^2\label{i3}
\end{eqnarray}
where $S^f=M_f\,M_f^\dagger$. This system can be solved and we find
\begin{eqnarray}
r^f&\approx& \frac{m^f_{3}}{\sqrt{m^f_{1}m^f_{2}}}\sqrt{\alpha^f} \label{ralpha}\,,\\
a^f&\approx& \frac{m^f_{2}}{m^f_{3}}\sqrt{\frac{m^f_{1}m^f_{2}}{\alpha^f}},\label{af}\,\\
b^f&\approx& \sqrt{\frac{m^f_{1}m^f_{2}}{\alpha^f}} \label{bf}.
\end{eqnarray}
in the limit $r^f\gg\alpha^f,1$ and $r^f\gg b^f/a^f$. These equations
are general in the sense that in the complex case, namely complex
Yukawa couplings and vevs, the invariants, Eqs. (\ref{i1})-(\ref{i3})
do not depend on the phases of the vevs $v_{u i}$. Indeed, the only
dependence on the relative phase of the Yukawa couplings, $y_{1}^f$ and
$y_2^2$ enters in the determinant, Eq. (\ref{i1}), and this is
negligible in the above limit.  From
Eqs.~(\ref{ralpha}),(\ref{af}),(\ref{bf}), one finds simple relations
for the second and third family masses, namely,
\begin{equation}
m^f_2 \approx a^fr^f, \ \ m^f_3 \approx b^fr^f, \ \ \frac{m^f_2}{m^f_3} \approx \frac{a^f}{b^f},
\end{equation}
from which we require $a^f \ll b^f$ in order to account for the second
and third family mass hierarchy. Moreover, since the charged leptons
and down-type quarks couple to the same Higgs and flavons, we have
\footnote{Note that this relation is natural in supersymmetric
  models~\cite{Morisi:2011pt,Bazzocchi:2012ve}. Here it follows from the $Z_N$ assignments, namely the fact
  that the same flavon $\varphi_d$ couple to down quarks and charged
  leptons.}
\begin{eqnarray}
r^l\,=\,r^d &\mbox{and}&\alpha^l\,=\,\alpha^d,
\end{eqnarray}
so that, from Eq.~(\ref{ralpha}), we obtain the mass
relation~\cite{Morisi:2011pt}
\begin{equation}
\frac{m_{\tau}}{\sqrt{m_{e}m_{\mu}}}\approx \frac{m_{b}}{\sqrt{m_{s}m_{d}}}  \,. 
\label{massrelationfinal}\end{equation}

\vskip3.mm Now we turn to the up-type quark sector.  From the
Lagrangian in Eq. (\ref{lagrangian}), the up quark mass matrix is
given by
\begin{equation}
M_u=
\left(
\begin{array}{ccc}
v_1^u & 0 & 0\\
0 & v_2^u & 0\\
0 & 0 & v_3^u
\end{array}
\right)\cdot
\left(
\begin{array}{ccc}
1& 1 &1\\
1 & \omega& \omega^2\\
1 & \omega^2& \omega\\
\end{array}
\right)\cdot
\left(
\begin{array}{ccc}
y^u_1 & 0 & 0\\
0 & y^u_{1^{\prime\prime}} & 0\\
0 & 0 & y^u_{1^{\prime}}
\end{array}
\right),\label{mu}
\end{equation}
In what follows we discuss the resulting structure of the quark mixing matrix.

\section{Quark mixing: the CKM matrix}
\label{sec4}


Recall that the down-type quark mass matrix takes the form in
Eq.~(\ref{Mf}) while the up-type quark mass matrix has just been given
in Eq.~(\ref{mu}). 
Out of these matrices one finds the matrices $M^u\cdot {M^u}^\dagger$
and $M^d\cdot {M^d}^\dagger$.  Their diagonalization results in two
unitary matrices $V^{u,d}$ for which one can obtain approximate
analytical expressions.  In the down sector, from $M^d\cdot
{M^d}^\dagger$, one finds,
\begin{eqnarray}
V^d_{12}&\approx&\sqrt{\frac{m_d}{m_s}}\frac{1}{\sqrt{\alpha^d}}  \\
V^d_{13}&\approx& \frac{m_s}{m_b^2}\sqrt{m_d m_s}\frac{1}{\sqrt{\alpha^d}}  \\
V^d_{23}&\approx& \frac{m_d m_s}{m_b^3} \frac{1}{\alpha^d}.
\end{eqnarray}
One sees that if $\alpha^d\sim\mathcal{O}(1)$ the down sector gives
about the Cabibbo angle in the $1-2$ plane while the mixings in the
$1-3$ and $2-3$ planes are negligible. 
On the other hand from the up quark sector one finds, approximately 
\begin{equation}
M_uM_u^\dagger\sim \left(
\begin{array}{ccc}
\lambda^8 & \lambda^6 & \lambda^4\\
\lambda^6 & \lambda^4 & \lambda^2\\
\lambda^4 & \lambda^2 & 1\\
\end{array}
\right)\label{msqlam}
\end{equation}
with coefficients of order one in front of the $(i,j)$ if at least one
of the Yukawa couplings is of order one\footnote{We note that the magnitude of the order-one parameters that enter implicitly in Eq.~(\ref{msqlam}) is relevant in order to fit the quark mass hierarchy. In particular note that the Yukawa couplings in Eq.~(\ref{mu}) have a hierarchical structure as determined from Eq.~(\ref{param}).}.  Thus for the up quark matrix
mixing factor $V^u$ we have that
\begin{eqnarray}
V^u_{23}&\approx& \lambda^2\,,\\
V^u_{13}&\approx& \lambda^4\,,\\
V^u_{12}&\approx& \lambda^2\,.
\end{eqnarray}
where we have assumed
\begin{eqnarray}
v_3^u: v_2^u: v_1^u & = & 1: \lambda^2 : \lambda^4.
\end{eqnarray}
The overall quark mixing matrix is given by the product
$${V^u}^\dagger\cdot V^d~.$$ One sees how the Cabibbo angle
arises from the down-type quark matrix mixing factor $V^d$, while the
$V_{ub}$ and $V_{cb}$ CKM mixing angles arise from the up quark matrix
mixing factor $V^u$.
Taking $\lambda\approx 0.2$ we obtain approximately the correct value
for the mixing angle.  However the order one parameters are relevant
in order to exactly determinate the quark mixing angles. In order to
obtain quantitative predictions for these we perform a global
numerical fit.  The experimental data used and the one~$\sigma$ error
bars are given in the second column of Table~\ref{input}, taking the quark masses (at the
scale of the $M_Z$) from~\cite{Bora:2012tx} and the quark mixing
angles from~\cite{Beringer:1900zz}.  The third column of Table~\ref{input} displays the
values predicted by our model when the values of its parameters are
those in equations~(\ref{param}).  

\vskip3.mm

We note that the phases of the up couplings $y_i^u$ can be reabsorbed
by transforming the right-handed fields $u_{R_i}$ while in the down
sector not all the phases can be removed.  For simplicity we assume
all the couplings to be real and we show how to make a fit of the
quark mixing parameters (including the complex phase) and masses. We
note that by taking $\omega$ to be the only phase in our
parametrization one can fit for the CKM phase, namely the Jarlskog
invariant~\footnote{However, this does not constitute a {\it
    geometrical origin} of the phase since the Yukawa couplings are
  complex.}.  
In table~\ref{input} we compare our theoretical predictions with the
current experimental values for the quark masses and CKM mixing
parameters.  The theoretical predictions are for the values:
\begin{equation}
\begin{array}{rcl}
y_1^u v_3^u  &=& -297393\,\textrm{MeV}, \\
y^u_{1^{\prime\prime}} v_3^u  &=& -15563\,\textrm{MeV}, \\
y_{1^{\prime}}^u v_3^u  &=& 277\,\textrm{MeV}, \\
v_2^u/v_3^u &=& 1.03 \lambda^2, \\
v_1^u/v_3^u &=& 2.14 \lambda^4, \\
\alpha_d &=& 1.58.\\
\end{array}
\label{param}
\end{equation}
where $\lambda=0.2$.

\begin{table}[t]
\begin{center}
\begin{tabular}{c|c|c}
\hline\hline Observable & Experimental value & Model prediction 
  \\ \hline
$m_{d}$~[MeV] & $2.9 \pm^{0.5}_{0.4}$ & $2.93$  \\
$m_{s}$~[MeV] & $57.7^{+16.8}_{-15.7}$ & $62$  \\
$m_{b}$~[MeV] & $2820^{+90}_{-40}$ & $2830$  \\
\\
$m_{u}$~[MeV] & $1.45^{+0.56}_{-0.45}$  & $1.63$  \\
$m_{c}$~[MeV] & $635 \pm 86$  & $640$  \\
$m_{t}$~[GeV] & $172.1 \pm 0.6\pm0.9$ & $172.1$  \\
\\
$|V_{us}|$ & $0.22534 \pm 0.00065$  & $0.2253$  \\
$|V_{ub}|$ & $0.00351^{+ 0.00015}_{-0.00014}$  & $0.00347$  \\
$|V_{cb}|$ & $0.0412^{+0.0011}_{-0.0005}$ & $0.0408$  \\
$J$ & $2.96^{+0.20}_{-0.16}$ & $2.93$ \\
\hline\hline
\end{tabular}
\end{center}
\caption{
  Experimental and predicted quark masses and mixing parameters from our fit.
  Quark masses (at the scale of the $M_Z$) have been taken 
  from~\cite{Bora:2012tx}, while  quark mixing angles have been taken 
  from~\cite{Beringer:1900zz}. The third column displays our predicted  
  values from Eqs.~(\ref{param}) which are in very good agreement 
  with the experimental data. }
\label{input}
\end{table}

\section{Neutrino masses and mixing}
\label{sec5}

As in Ref.~\cite{Morisi:2011pt} here we consider an effective way to
generate neutrino masses by {\it \`{a} la Weinberg} by upgrading the
standard dimension-five operator to the flavon case, making it
dimension-six, that is~\footnote{It is easy to formulate
  type-II~\cite{Schechter:1980gr,Lazarides:1980nt,mohapatra:1981yp} or
  inverse~\cite{mohapatra:1986bd} seesaw variants of the
  model. However for simplicity here we keep the effective
  description presented above.}
\begin{equation}
\frac{y_\varphi^\nu}{\Lambda^2}LLHH\varphi_\nu+\frac{y_\xi^\nu}{\Lambda^2}LLHH\xi_\nu\,.
\label{nuops}
\end{equation}
After electroweak symmetry breaking it gives to the following
Majorana neutrino mass matrix
\begin{equation}
m_\nu=\left(
\begin{array}{ccc}
d & a & b\\
 & d & c\\
 &  & d
\end{array}
\right)
\end{equation}
where $a=y_\varphi^\nu / \Lambda^2 v_H^2 \vev{\varphi_{\nu\,3}}$,
$b=y_\varphi^\nu / \Lambda^2 v_H^2 \vev{\varphi_{\nu\,2}}$,
$c=y_\varphi^\nu / \Lambda^2 v_H^2 \vev{\varphi_{\nu\,1}}$ and
$d=y_\xi^\nu / \Lambda^2 v_H^2 \vev{\xi_\nu}$.  Note that, unlike the
charged fermion case, only the symmetric contractions are allowed from
the first operator in Eq.~(\ref{nuops}).  We remark that these parameters
in the neutrino sector are unrelated to those in the charged fermion
sector. 

\begin{figure}[h!]
\begin{center}
\includegraphics[trim=5cm 14.5cm 3cm 5.5cm,scale=.8]{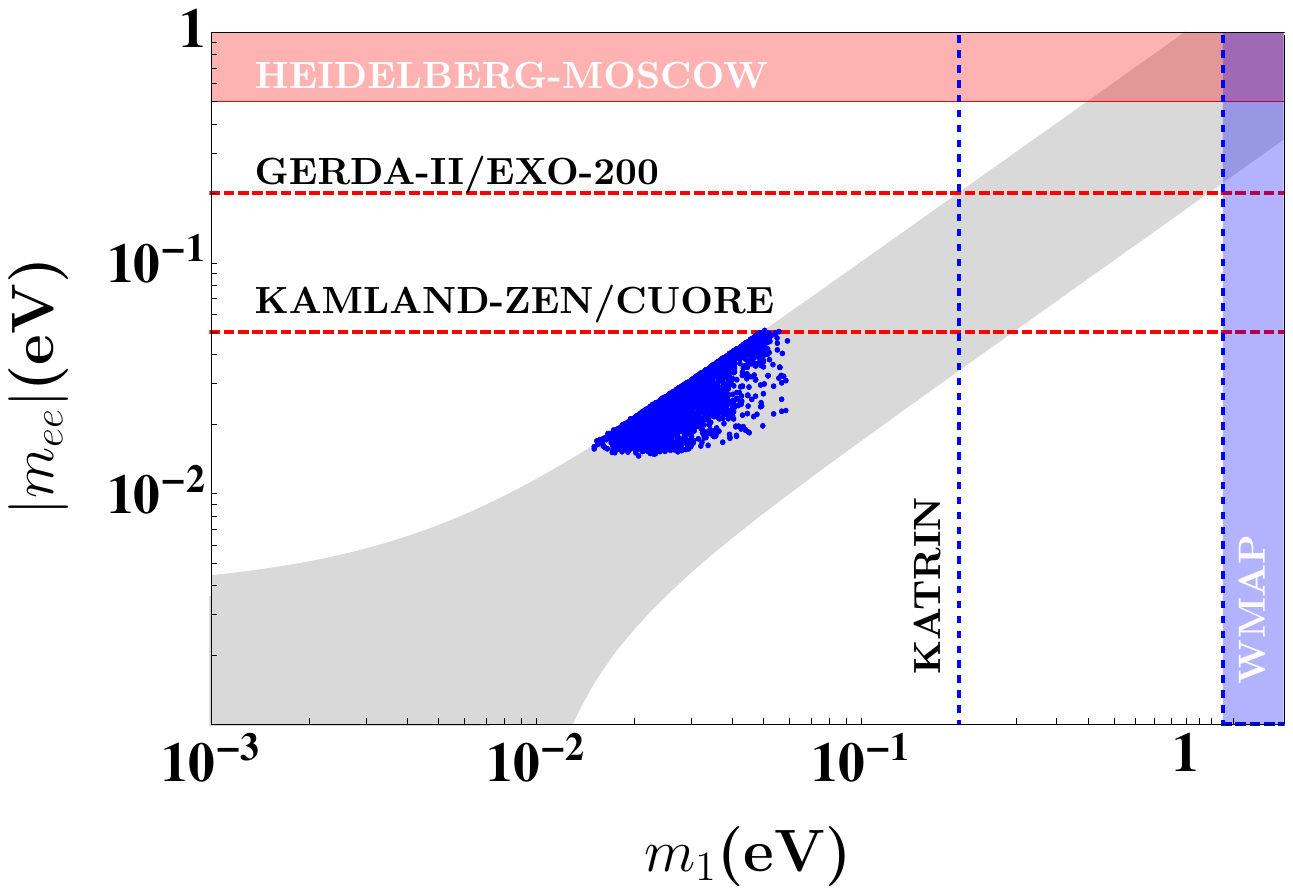} 
\caption{Effective neutrino mass parameter $m_{ee}$ as function of the
  lightest neutrino mass. The gray shaded regions correspond to the
  flavour-generic normal hierarchy neutrino spectra. The blue points
  correspond to the prediction of our $A_4$ model. For comparison we give the current and future sensitivities for $m_{ee}$~\cite{Schwingenheuer:2012jt,Rodejohann:2011mu} and $m_\nu$~\cite{Osipowicz:2001sq,Komatsu:2010fb}, respectively.}
\label{fig2}\end{center}
\end{figure}

\begin{figure}[h!]
\begin{center}
 \includegraphics[width=7cm]{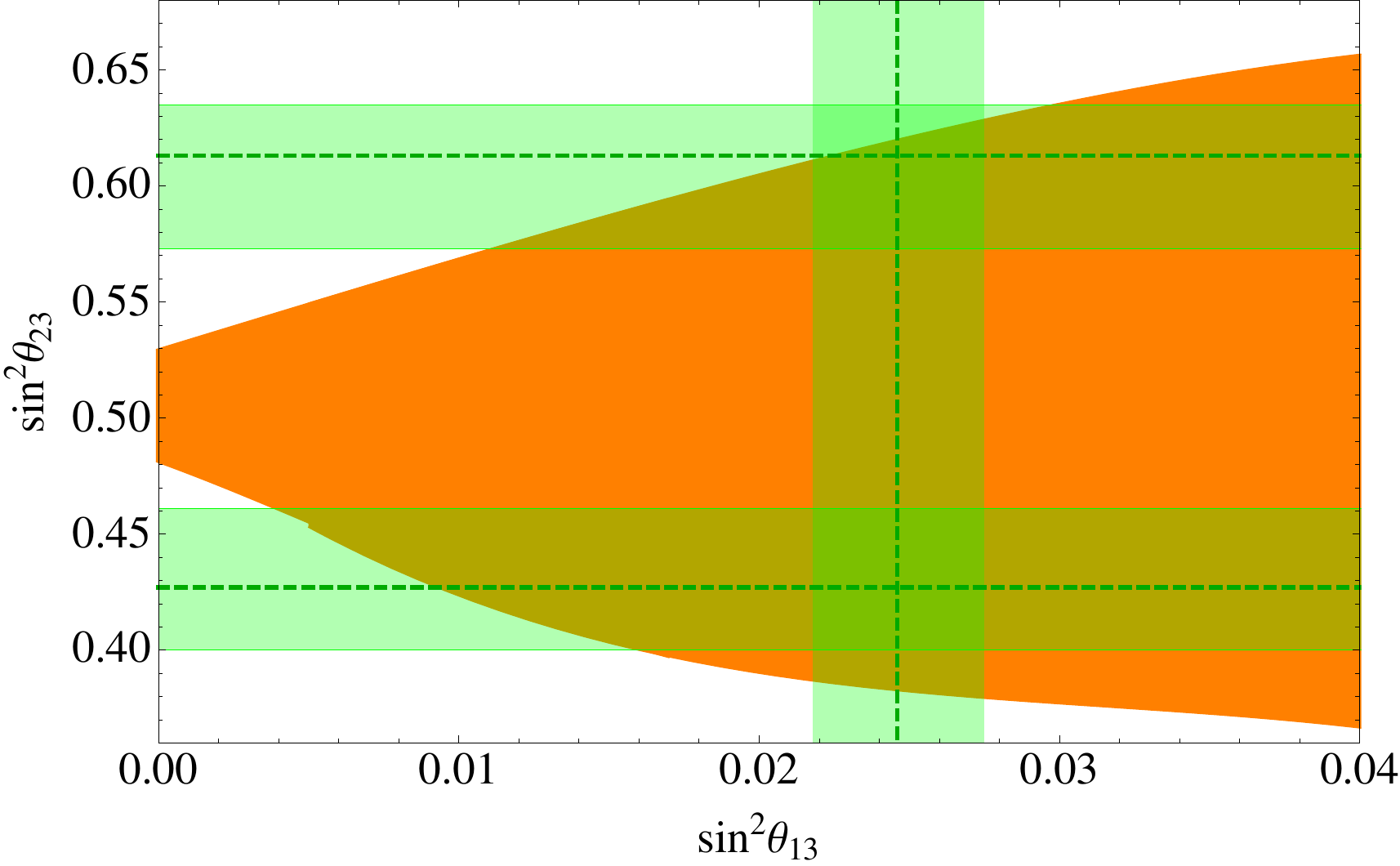}  
 \includegraphics[width=7cm]{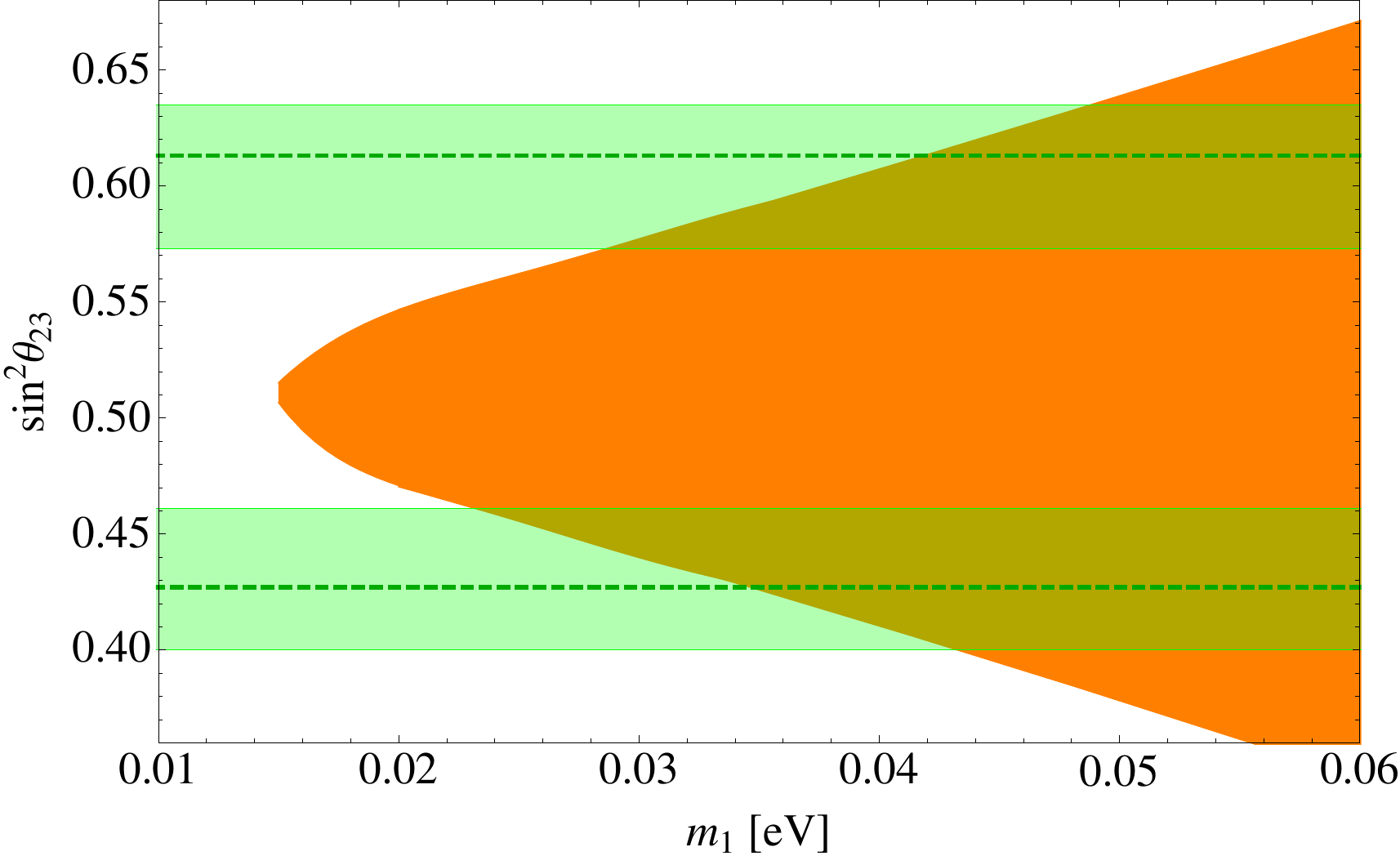}
 \caption{Correlations between the atmospheric angle, and the reactor
   angle (left panel) and the lightest neutrino mass (right panel).
   Straight bands are the currently allowed 1~$\sigma$ bands of the
   oscillation angles, taken from Ref.~\cite{Tortola:2012te}.}
\label{fig1}\end{center}
\end{figure}

Taking the limit where $d=0$ the neutrino mass matrix has the well
known Zee-Wolfenstein structure, which cannot reproduce the current
neutrino oscillation data~\cite{Tortola:2012te}, since it gives to
Bi-maximal mixing \cite{Jarlskog:1998uf}. The addition of the unit
matrix contribution proportional to $d$ introduces deviations from
maximal atmospheric mixing proportional to $b$ and also introduces a
non-zero reactor angle $\theta_{13}\sim (a-b)/(2d)$, while the solar
angle is approximately given by $\tan 2\theta_{12}\sim 2(a+b)/d$,
which reduces to maximal solar mixing in the Zee-Wolfenstein limit $d
\to 0$. One finds a strict correlation between the neutrinoless double
beta decay rate and the magnitude of the parameter $d$. In fact, as
seen in Fig.~\ref{fig2} we find a (weak) lower bound for the
neutrinoless double beta decay rate, despite the fact that the model
has a normal hierarchy neutrino spectrum, this follows from the
presence of the flavour symmetry~\footnote{Similar examples of $A_4$
  models with similar features can be found in~\cite{Hirsch:2005mc}.}.
In our numerical scan we also obtain a restricted set of neutrino
oscillation angles. For example the curved-shaped region in the left
panel of Fig.~\ref{fig1} (orange in color version) corresponds to the
``predicted'' atmospheric angle consistent with the currently allowed
values of the solar angle at 3~$\sigma$, from
Ref.~\cite{Tortola:2012te}. For comparison we display also the
1~$\sigma$ bands for $\sin^2\theta_{23}$ and $\sin^2\theta_{13}$. In
the right panel in Fig.~\ref{fig1} we re-express our
$\sin^2\theta_{23}$ ``prediction'' in terms of the lightest neutrino
mass $m_1$, again keeping undisplayed oscillation parameters at
3~$\sigma$.
The existence of the above restrictions reflects the fact that, as a
result of the flavour symmetry, we have in total less parameters than
observables to describe.

\section{Conclusions}
\label{sec6}

We have proposed a realistic $A_4$ extension of the Standard Model
leading to the quark-lepton mass relation given in
Eq.~(\ref{massrelation}).  This successful and nearly renormalization
invariant mass relation generalizes the Georgi-Jarlskog formula and
arises outside the context of unification.  Quark masses, mixing
angles and CP violation are properly accounted for, while in the
neutrino sector we obtain a generalized Zee-Wolfenstein mass matrix
giving an acceptable reactor angle along with a deviation of the
atmospheric and solar angles from their bi-maximal values.  As seen in
Fig.~\ref{fig2} the neutrinoless double beta decay rate correlates
sharply with this deviation parameter, with a minimum allowed value
despite the fact that the model has a normal hierarchy neutrino mass
spectrum. Moreover we find that the atmospheric angle correlates with
the lightest neutrino mass (right panel in Fig.~\ref{fig1}) and with
the reactor angle (left panel).  From the theory point of view the
model treats all fermion masses effectively, as arising from
non-renormalizable Yukawa-like terms.

\appendix
\section{The product rules for the $A_4$ group}
\label{a4group}

Here we adopt the $SO(3)$ basis for the generators of the $A_4$ group,
which can be written as $S$ and $T$ with $S^2=T^3=(ST)^3=\mathcal{I}$.
$A_4$ has four irreducible representations, three singlets
$1,~1^\prime$ and $1^{\prime \prime}$ and one triplet.  In the basis
where $S$ is real diagonal,
\begin{equation}\label{eq:ST}
S=\left(
\begin{array}{ccc}
1&0&0\\
0&-1&0\\
0&0&-1\\
\end{array}
\right)\,;\quad
T=\left(
\begin{array}{ccc}
0&1&0\\
0&0&1\\
1&0&0\\
\end{array}
\right)\,;
\end{equation}
The products of singlets are:
\begin{equation}\begin{array}{llll}
1\otimes1=1&1^\prime\otimes1^{\prime\prime}=1&1^{\prime}\otimes1^{\prime}=1^{\prime\prime}
&1^{\prime\prime}\otimes1^{\prime\prime}=1^\prime

\end{array}
\end{equation}
one has the following triplet multiplication rules, 
\begin{equation}\label{pr}
\begin{array}{lll}
(ab)_1&=&a_1b_1+a_2b_2+a_3b_3\,;\\
(ab)_{1'}&=&a_1b_1+\omega a_2b_2+\omega^2a_3b_3\,;\\
(ab)_{1''}&=&a_1b_1+\omega^2 a_2b_2+\omega a_3b_3\,;\\
(ab)_{3_1}&=&(a_2b_3,a_3b_1,a_1b_2)\,;\\
(ab)_{3_2}&=&(a_3b_2,a_1b_3,a_2b_1)\,,
\end{array}
\end{equation}
where $\omega^3=1$, $a=(a_1,a_2,a_3)$ and $b=(b_1,b_2,b_3)$.

\section*{Acknowledgments}

Work supported by the Spanish MINECO under grants
FPA2011-22975 and MULTIDARK CSD2009-00064 (Consolider-Ingenio 2010
Programme), by Prometeo/2009/091 (Generalitat Valenciana), and by the
EU ITN UNILHC PITN-GA-2009-237920. S.F.K. also acknowledges partial support 
from the STFC Consolidated ST/J000396/1 and EU ITN INVISIBLES 289442. S. M. 
acknowledges support by the DFG grant WI 2639/4-1.


\end{document}